\documentclass{epl}

\begin{document}

\title{The Metal-Insulator Transition of the Magn\'{e}li phase 
       $ {\rm \bf V_4O_7} $: Implications for $ {\rm \bf V_2O_3} $} 

\shorttitle{The Metal-Insulator Transition of $ {\rm V_4O_7} $}

\author{U.\ Schwingenschl\"ogl
        \and V.\ Eyert 
        \thanks{E-mail: \email{Volker.Eyert@physik.uni-augsburg.de}}
	\and U.\ Eckern}
\shortauthor{U.\ Schwingenschl\"ogl \etal}

\institute{Institut f\"ur Physik, Universit\"at Augsburg, 
           86135 Augsburg, Germany}


\maketitle

\begin{abstract}
The metal-insulator transition (MIT) of the Magn\'{e}li phase 
$ {\rm V_4O_7} $ is studied by means of electronic structure calculations 
using the augmented spherical wave method. The calculations are based on 
density functional theory and the local density approximation. Changes 
of the electronic structure at the MIT are discussed in relation to the 
structural transformations occuring simultaneously. The analysis is based 
on a unified point of view of the crystal structures of all Magn\'{e}li 
phase compounds $ {\rm V_nO_{2n-1}} $ ($3 \leq n \leq 9 $) as well as of 
$ {\rm VO_2} $ and $ {\rm V_2O_3} $. This allows to group the electronic 
bands into states behaving similar to the dioxide or the sesquioxide. 
In addition, the relationship between the structural and electronic 
properties near the MIT of these oxides can be studied on an equal 
footing. For $ {\rm V_4O_7} $, a strong influence of metal-metal bonding 
across octahedral faces is found for states both parallel and perpendicular 
to the hexagonal $ c_{\rm hex} $ axis of $ {\rm V_2O_3} $. Furthermore, the 
structural changes at the MIT cause localization of those states, which 
mediate in-plane metal-metal bonding via octahedral edges. This band 
narrowing opens the way to an increased influence of electronic correlations, 
which are regarded as playing a key role for the MIT of $ {\rm V_2O_3} $. 
\end{abstract}

The vanadium oxides have been attracting a lot of interest for many years, 
in particular due to their metal-insulator transitions (MIT). Special 
focus has been on the prototypical compounds ${\rm VO_2}$ and ${\rm V_2O_3}$, 
which were studied quite extensively \cite{goodenough71,brueckner83,imada98}. 
However, much dispute remains as concerns the origin of the phase transitions. 
While it is agreed that these arise from a delicate interplay of 
electron-phonon coupling and electronic correlations, the relative 
importance of these two mechanisms is still under controversal discussion.

At a temperature of 340\,K, ${\rm VO_2}$ ($3d^1$) undergoes an MIT as well 
as a simultaneous structural transformation from the rutile to a monoclinic 
structure. This combined phase transition can be understood from electronic 
structure calculations, which point to a Peierls instability of the 
one-dimensional $d_{\parallel}$ ($d_{x^2-y^2}$) band in an embedding
background of the remaining V $3d$ $t_{2g}$ states \cite{eyert02b}. 
Nevertheless, 
the optical band gap of the insulating phase was missed by the calculations 
due to the shortcomings of the local density approximation (LDA). However, 
the interpretation in terms of an embedded Peierls instability was supported 
by calculations for ${\rm MoO_2}$ and ${\rm NbO_2}$ \cite{eyert00a,eyert02a}. 
The latter material likewise undergoes an MIT, which is accompanied by 
structural distortions very similar to those of ${\rm VO_2}$. Since within 
the LDA a finite optical band gap was obtained for the niobium compound but 
slightly missed for ${\rm VO_2}$, it was concluded that electronic 
correlations play an albeit small role in the $ 3d $ system 
\cite{eyert02a,eyert02b}. 

Stoichiometric ${\rm V_2O_3}$ ($3d^2$) shows an MIT at 168\,K under  
ambient pressure, leading from a paramagnetic metallic (PM) to an 
antiferromagnetic insulating (AFI) phase. Again, there is a simultaneous 
change of the crystal structure from corundum type to a monoclinic lattice. 
On doping with small amounts of Al or Cr, ${\rm V_2O_3}$ displays a 
paramagnetic insulating (PI) phase, which still has the corundum structure. 
However, EXAFS experiments revealed local structural distortions in the PI 
phase, which are identical to those characterizing the AFI phase but lack 
long-range order \cite{pfalzer02}. The phase transitions of ${\rm V_2O_3}$ 
are commonly regarded as of the Mott- or Mott-Hubbard type, driven 
by electronic correlations of the V $3d$ states \cite{castellani78}. While 
LDA calculations for all three phases show only a minor response of the 
electronic properties to the crystal structure changes occuring at the 
transitions, LDA+U calculations reproduced the insulating band gap of the 
AFI phase \cite{mattheiss94,ezhov99}. In contrast, a correct description 
of the PM-PI phase transition was achieved only recently by a combination 
of LDA calculations with the dynamical mean-field theory (DMFT) \cite{held01}. 
In addition, this new approach reproduced the vanadium $ S = 1 $ state 
called for by results from polarized x-ray spectroscopy experiments 
\cite{park00}. 

In order to arrive at a deeper understanding of the MIT of the afore 
mentioned vanadium oxides, an identification of the relevant electronic 
states as well as their response to changes of the crystal structures 
is highly desirable. In this paper we address this issue by studying 
the relation between structural and electronic properties in the 
broader class of the vanadium Magn\'eli phases. These compounds form a 
homologous series $ {\rm V_nO_{2n-1}} $ $ (3\leq n\leq9) $ with crystal 
structures built from characteristic dioxide-like and sesquioxide-like 
regions. Thus they may be regarded as intermediate between the structures 
of ${\rm VO_2}$ ($ n \to \infty $) and ${\rm V_2O_3}$ ($ n = 2 $), 
allowing for a deeper insight into the crossover from the dioxide 
to the sesquioxide. Like ${\rm VO_2}$ and ${\rm V_2O_3}$, all Magn\'eli 
phases except for ${\rm V_7O_{13}}$ show MIT, which are accompanied by 
structural transformations. Furthermore, they exhibit long-range 
antiferromagnetic order at N\'eel temperatures, which, except 
for $ {\rm VO_2} $ and $ {\rm V_2O_3} $, are much lower than the 
temperatures connected with the MIT \cite{kachi73}. From this it 
was concluded that the insulating state as well as the MIT are not 
related to the antiferromagnetic order and, hence, should be studied 
independently. X-ray refinements suggested that the MIT 
goes along with charge ordering separating $ {\rm V^{3+}} $ and 
$ {\rm V^{4+}} $ sites \cite{marezio72}. NMR measurements confirmed 
this result in general  but found the amount of charge disproportionation 
to be much smaller than one \cite{gossard74}. While our previous study 
on ${\rm V_6O_{11}}$ gave strong indications for the predominant 
influence of electron-lattice interaction in the dioxide-like regions 
and of electronic correlations in the sesquioxide-like regions of the 
crystal, respectively \cite{us03}, the present work concentrates on the 
sesquioxide-related member $ {\rm V_4O_7} $ and, hence, allows to draw 
conclusions about the MIT of ${\rm V_2O_3}$. 

\begin{figure}
\oneimage[width=144mm]{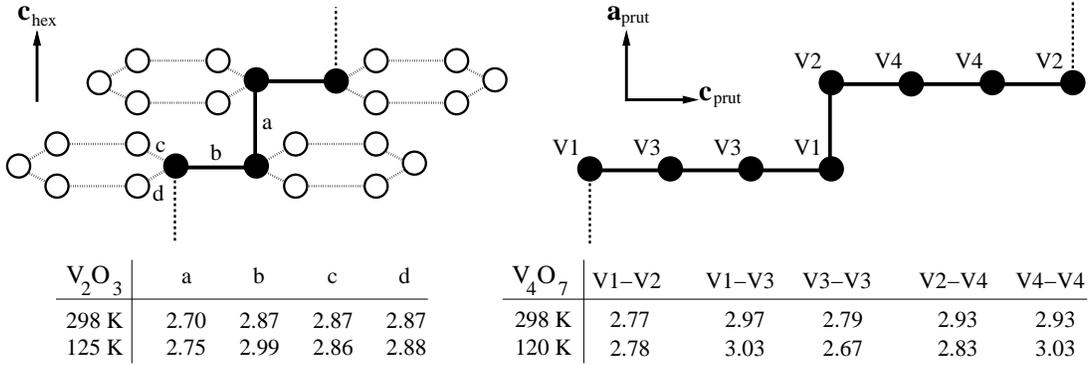}
\caption{Building blocks of the crystal structures of ${\rm V_2O_3}$ 
         (left side) and ${\rm V_4O_7}$ (right side). Only vanadium 
	 atoms are shown. The tables list measured V-V distances for 
	 both the metallic and the insulating phase.}
\label{fig1}
\end{figure}

As the general formula
$ {\rm V}_n{\rm O}_{2n-1} = {\rm V_2O_3} + (n-2) {\rm VO_2} $ suggests,
the crystal structures of the Magn\'{e}li phases are usually viewed as
rutile-type slabs of infinite extension and different thickness, which 
are separated by shear planes with a corundum-like atomic arrangement
\cite{brueckner83,magneli48,andersson63}. Recently, we have proposed 
an alternative unified representation of the crystal structures of all 
Magn\'eli phases in terms of a regular 3D network of oxygen octahedra, 
which are partially filled by vanadium atoms \cite{us03}. The filled 
octahedra form chains of length $n$ parallel to the pseudorutile 
$c_{\rm prut}$ axis, followed by $n-1$ empty sites. While in ${\rm VO_2}$ 
these chains have infinite length, they comprise just two filled octahedra 
in ${\rm V_2O_3}$. The situation is sketched in Fig.\ \ref{fig1} 
for ${\rm V_2O_3}$ and ${\rm V_4O_7}$. Note that only the vanadium atoms 
are depicted for simplicity. Along the $a_{\rm prut}$ axis, which is 
identical to the hexagonal axis $c_{\rm hex}$ of ${\rm V_2O_3}$, vanadium 
and oxygen layers alternate. In the Magn\'eli phases, two different types 
of vanadium layers are distinguished, which likewise alternate along 
$a_{\rm prut}$. In ${\rm V_4O_7}$ the layers comprise the atoms V1/V3 and 
V2/V4, respectively. Due to the alternation of the layers and relative 
shifts of the four-atom chains the end atoms V1 and V2 are found on top of 
each other, see Fig.\ \ref{fig1}. While neighbouring octahedra share faces 
along $a_{\rm prut}$ and $b_{\rm prut}$, metal-metal bonding along all 
other directions within the layers is via edges (see Ref.\ \cite{eyert02b} 
for further details). As a consequence, the 
atomic arrangement near the chain ends is ${\rm V_2O_3}$-like, whereas the 
chain centers correspond to the rutile-type regions. By virtue of the just 
sketched representation of the crystal structures it is possible to refer 
the symmetry components of the V $3d$ orbitals of all compounds to a common 
local coordinate system \cite{eyert02b}. In this system the $ z $- and 
$ x $-axes of the local coordinates are parallel to the apical axis of the 
local octahedron and the pseudorutile $c_{\rm prut}$-axis, respectively. 

Our LDA calculations are based on the scalar-relativistic augmented spherical 
wave (ASW) method \cite{williams79,eyert00b}. Crystallographic data as given 
by Hodeau and Marezio \cite{hodeau78} were used. In order to represent the 
correct shape of the crystal potential in the large voids of the open crystal 
structure of $ {\rm V_4O_7} $, additional augmentation spheres were inserted.  
Optimal augmentation sphere positions as well as radii of all spheres were 
automatically generated by the sphere geometry optimization (SGO) algorithm 
\cite{eyert98b}. The basis sets comprised V $4s$, V $4p$, V $3d$, O $2s$, 
and O $2p$ orbitals as well as states of the additional augmentation spheres. 
Brillouin zone integrations were performed using an increasing number of 
${\bf k}$ points, ranging from 108 to 2048 points within the irreducible 
wedge.

\begin{figure}
\oneimage[width=144mm]{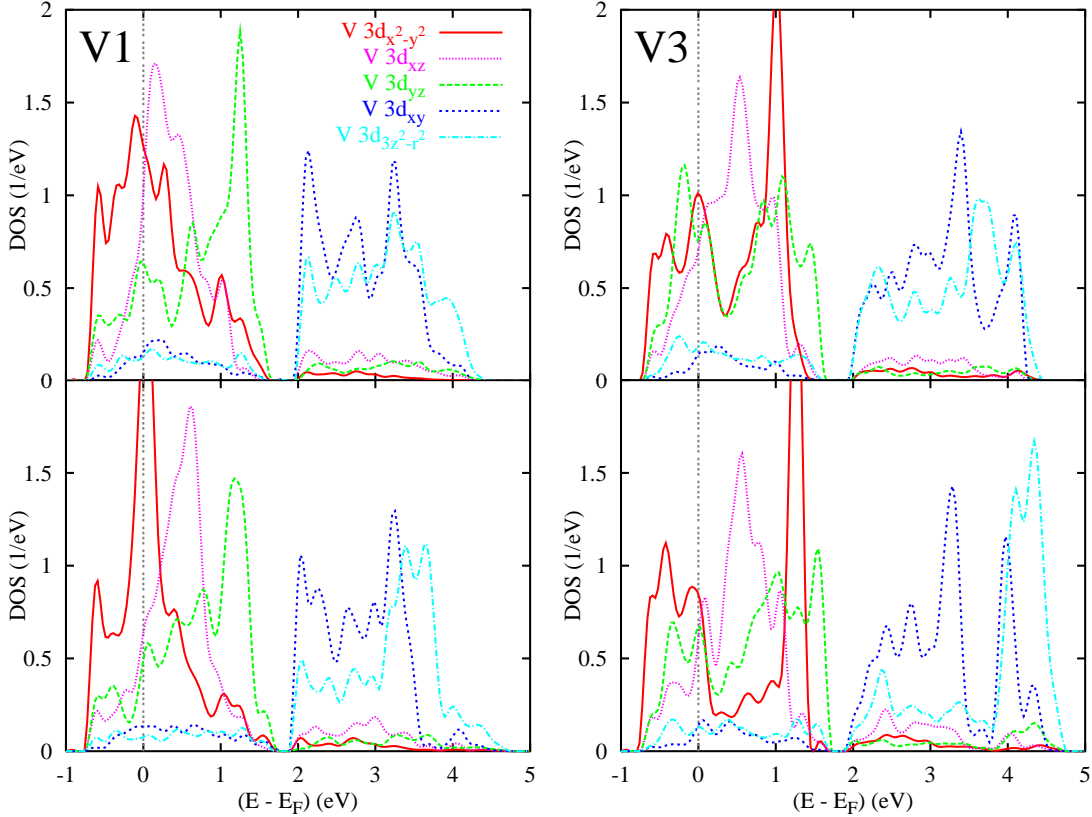}
\caption{Partial V1 and V3 $3d$ DOS (per atom) of high- (top) and 
         low-temperature (bottom) ${\rm V_4O_7}$.}
\label{fig2}
\end{figure}

\begin{figure}
\oneimage[width=144mm]{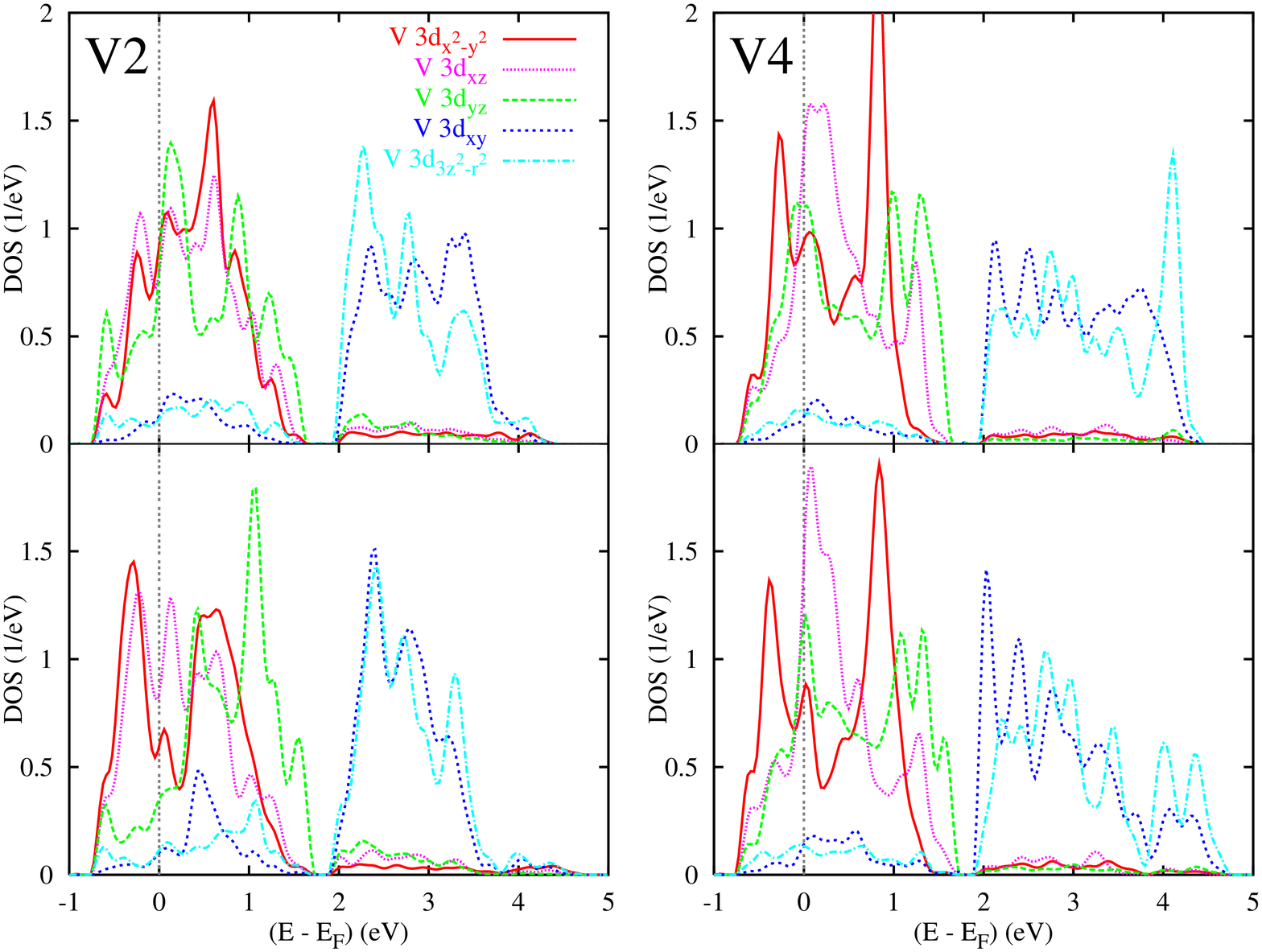}
\caption{Partial V2 and V4 $3d$ DOS (per atom) of high- (top) and 
         low-temperature (bottom) ${\rm V_4O_7}$.}
\label{fig3}
\end{figure}

Partial V $3d$ densities of states (DOS) resulting from the calculations 
for the crystal structures of both phases of ${\rm V_4O_7}$ are displayed 
in Figs.\ \ref{fig2} 
and \ref{fig3}. 
In the present paper, we concentrate on the results for the V1/V3 chains. 
This is motivated by close similarity of the changes in the local 
surrounding of atom V1 as coming with the structural transformations at 
the MIT to those known from $ {\rm V_2O_3} $. According to 
Fig.\ \ref{fig1}, the V1-V2 and V1-V3 distances both increase 
at the MIT just as the distances a and b in the sesquioxide at the PM-AFI 
transition. Since in the latter compound all vanadium 
atoms are still crystallographically identical, a complete symmetry analysis 
of the electronic states could not be performed. However, this is possible 
for $ {\rm V_4O_7} $. As we have demonstrated in our work on 
$ {\rm V_6O_{11}} $ the local atomic arrangements can be directly related 
to the corresponding electronic properties at these sites \cite{us03}. 
Hence, the V1 site of ${\rm V_4O_7}$ is particularly suited to study the 
influence of the structural changes on the MIT of ${\rm V_2O_3}$. 

The gross features of the DOS from Figs.\ \ref{fig2} and \ref{fig3} are 
very similar to those known from ${\rm VO_2}$ and ${\rm V_2O_3}$. As a 
consequence of the crystal field splitting of the V $3d$ states due to 
the surrounding oxygen octahedra one finds two groups of bands in the 
energy range displayed. Note that all orbitals are referred to the 
above mentioned local coordinate systems \cite{eyert02b}. 
V $3d$ $t_{2g}$ and $e_g$ states dominate in the energy regions from 
about $-0.6$ to $1.8$\,eV and $2.0$ to $4.7$\,eV, respectively. 
O $2p$ states, which give larger contributions in the energy region 
from $-8.3$ to $-3.0$\,eV, are not included. Oxygen and vanadium 
contributions in the energy regions where the respective other orbital 
dominates are less than 10\% but still indicative of covalent bonding. 
Obviously, the densities of states obtained for the low-temperature phase 
miss an insulating band gap. As mentioned above, this effect is well 
known from $ {\rm V_2O_3} $ and mirrors the shortcomings of the LDA. However, 
since in the present context we are primarily interested in understanding 
the relationship between changes of the crystal and the electronic structure 
at the MIT, these limitations do not prohibit the following considerations. 
Our low-temperature phase results show a charge disproportionation at the 
vanadium sites of the order of 0.2 electrons, which is in close agreement 
with the NMR data \cite{gossard74}. 

Taking a closer look at the V $ 3d $ $ t_{2g} $ partial DOS we find for 
the $ d_{x^2-y^2} $ DOS a similar behaviour as already observed in 
$ {\rm VO_2} $ \cite{eyert02b}. Due to strong $ \sigma $-type metal-metal 
bonding parallel to the vanadium chains the chain-center atoms display a 
distinct splitting of these states. At low temperatures the chains behave 
differently. Strong V-V dimerization leads even to an enhanced 
bonding-antibonding splitting of the V3 $ d_{x^2-y^2} $ states. In addition, 
it enforces a sharpening, hence, a stronger localization of the 
$ d_{x^2-y^2} $ states on the V1 atoms, which have only one neighbour 
along the chains. In contrast, the V4-V4 distance hardly changes, whereas 
the V2-V4 distance shrinks, this causing an increased bonding-antibonding 
splitting of the V2 $ d_{x^2-y^2} $ states. 

Changes of the vanadium $ d_{xz} $ partial DOS at the transition are less 
significant and consist mainly of an energetical up- and downshift of these 
states as observed for atom V1 and V2, respectively. These shifts are due 
to increased and decreased lateral (perpendicular to $ c_{\rm prut} $) 
vanadium atom displacements away from the centers of gravity of the 
surrounding oxygen octahedra and, hence, derive from the changes of the 
V $ 3d $-O $ 2p $ overlap. They are thus well understood in terms of the 
antiferroelectric zigzag-like displacement already observed for vanadium 
dioxide \cite{eyert02b}. However, note that in the latter compound finite 
lateral displacements occur only in the low-temperature phase, whereas 
$ {\rm V_4O_7} $ displays them in both phases. Interestingly, the lateral 
displacement is much larger for the chain-end atoms as compared to atoms 
V3 and V4. This is due to a slight rotation of the chains away from the 
pseudorutile axis $ c_{\rm prut} $, which has been observed for all 
Magn\'{e}li phases \cite{us03}. In passing, we mention that metal-metal 
overlap is less important for the $ d_{xz} $ orbitals. 

The situation is more complicated for the V $ d_{yz} $ orbitals, which 
are strongly influenced by different kinds of hybridizations. In 
$ {\rm VO_2} $, these orbitals take part in metal-metal overlap across 
octahedral faces along both $ a_{\rm rut} $ and $ b_{\rm rut} $ 
\cite{eyert02b}. In addition, the antiferroelectric displacement of the 
vanadium atoms leads to increased $ d $-$ p $ overlap causing an upshift 
of the antibonding, $ d $-dominated states. In $ {\rm V_4O_7} $, the atoms 
V3 and V4 display neither considerable antiferroelectric displacements  
nor metal-metal overlap along $ a_{\rm prut} $, since there are no vanadium 
neighbours along this direction. As a consequence, the pronounced 
double-peak structure of the V3 and V4 $ d_{yz} $ partial DOS is indicative 
of bonding-antibonding splitting due to V1-V3 and V2-V4 overlap, respectively,  
across the shared octahedral faces parallel to $ b_{\rm prut} $. The 
chain-end atoms, in contrast, apart from participating in the V1-V3 and V2-V4 
overlap parallel to $ b_{\rm prut} $, take part in V1-V2 bonding along  
$ a_{\rm prut} $. The $ d_{yz} $ partial DOS of atoms V1 and V2 thus display 
features reminiscent of the corresponding partial DOS of atoms V2/V3 and 
V1/V4, respectively. In particular, we observe the close resemblance of 
the $ d_{yz} $ partial DOS of these two atoms in the low-temperature 
phase. Being subject to two different types of overlap, the $ d_{yz} $ 
orbitals of V1 and V2 undergo two intertwining bonding-antibonding 
splittings leading to the complicated triangular-like shape of the 
corresponding partial DOS similar to that known from the V $ 3d $ $ a_{1g} $ 
states of $ {\rm V_2O_3} $ \cite{held01}. In addition, their form is 
affected by the above mentioned zigzag-like displacement of the chain-end 
atoms causing increased metal-oxygen overlap at these sites. As a 
consequence, the centers of gravity of the V1 and V2 $ d_{yz} $ partial 
DOS are shifted to higher energies. Finally, the changes of the V2 
$ d_{yz} $ partial DOS, in particular the closer resemblance of this DOS 
to the respective curve of atom V1 in the low-temperature phase, can be 
traced back to the reduced relative displacement of these atoms parallel 
to $ c_{\rm prut} $. This causes an improved alignment of the V1-V2 bond 
parallel to $ a_{\rm prut} $ and, hence, an increased overlap along this 
latter direction. 

To summarize our results, we find the electronic properties of 
$ {\rm V_4O_7} $ to be strongly influenced by the local environment of 
each atom. While, in general, the overlap of O $ 2p $ and V $ 3d $ states 
places the V $ 3d $ $ t_{2g} $ states near the Fermi energy, the detailed 
electronic structure of the latter are subject to the local metal-metal 
coordination. In particular, we have found a strong bonding-antibonding 
splitting of the $ d_{x^2-y^2} $ states of the chain-center atoms and, 
in the  low-temperature structure, also for atom V2. This is in close 
analogy to the behaviour of these states in $ {\rm VO_2} $, where it is 
likewise enforced by metal-metal dimerization. In contrast, the 
V1 $ d_{x^2-y^2} $ states display a distinct sharpening since they do 
not take part in V-V dimerizations along the chains. Hence, they are more 
susceptible to electronic correlations.  A similar behaviour 
should be observed for $ {\rm V_2O_3} $, where the corresponding distance 
(labelled b in Fig.\ \ref{fig1}) likewise increases. 

For the $ d_{xz} $ and $ d_{yz} $ states we observe a distinct response 
to the displacement of the vanadium atoms perpendicular to $ c_{\rm prut} $ 
relative to the centers of their respective octahedra, leading to energetical 
up- and downshift of these states. However, the $ d_{yz} $ partial DOS of 
all atoms are mainly affected by the metal-metal bonding across the shared 
octahedral faces. As a consequence, they show bonding-antibonding 
splitting as seen in the $ d_{yz} $ partial DOS of the chain-center atoms, 
which are subject to bonding only parallel to $ b_{\rm prut} $. The 
situation is more complicated for atoms V1 and V2, which take part in the 
just mentioned V1-V3 and V2-V4 bonding parallel to $ b_{\rm prut} $ as well 
as in V1-V2 bonding along $ a_{\rm prut} $. They end up with a $ d_{yz} $ 
partial DOS similar to that of the $ a_{1g} $ states of $ {\rm V_2O_3} $. 
Note that the latter states mediate metal-metal bonding parallel to 
$ a_{\rm prut} $ ($ \, \hat{=} \, c_{\rm hex} $). 

In conclusion, similar to the situation in other Magn\'{e}li phases 
\cite{us03}, the electronic structure of $ {\rm V_4O_7} $ arises as a 
mixture of features characteristic of either the dioxide or the 
sesquioxide. Note that the local environment of the chain-end atom V1 
bears a close resemblance to the situation in $ {\rm V_2O_3} $. In both 
compounds this atom shows an increasing separation from its vanadium 
neighbours at the transition.  From our results for $ {\rm V_4O_7} $ we 
are thus able to draw the following conclusions for the sesquioxide: 
First, while the metal-metal bonding across the octahedral faces parallel 
to $ a_{\rm prut} $ seems to have the largest effect on the $ d_{yz} $, 
hence, the $ a_{1g} $ states of $ {\rm V_2O_3} $, $ d $-$ d $ hybridization 
across the faces parallel to $ b_{\rm prut} $ is rather large. Instead, 
it causes strong splitting of the electronic states. 
Second, due to the reduced metal-metal bonding across the octahedral edges 
parallel to $ c_{\rm prut} $ the V $ d_{x^2-y^2} $ states should undergo 
increased localization at the transition also in $ {\rm V_2O_3} $. This 
conclusion supports previous work by Dernier, who concluded from a 
comparison of pure and doped $ {\rm V_2O_3} $ as well as $ {\rm Cr_2O_3} $ 
that the metallic properties are intimately connected with the V--V 
hybridization across the shared octahedral edges rather than with hopping 
processes within the vanadium pairs along the hexagonal $ c $ axis 
\cite{dernier70a}.
From this point of view the MIT thus arises from the crystal structure 
changes occuring at the transition. Via strong electron-phonon coupling 
these distortions translate into a narrowing of the bands perpendicular to 
$ c_{\rm hex} $ and, eventually, allow for an increased influence of 
electronic correlations in the insulating phase. From the presence of 
local octahedral distortions in the PI phase as reported by Pfalzer 
{\it et al.}\ \cite{pfalzer02}, we expect this scenario to hold for both 
the PM-AFI and the PM-PI phase transition.

\begin{acknowledgments}
Fruitful discussions with S.\ Horn, S.\ Klimm, and P.\ Pfalzer are gratefully
acknowledged. This work was supported by the Deutsche Forschungsgemeinschaft 
through SFB 484. 
\end{acknowledgments}

\end{document}